\documentclass[10pt, final, conference, letterpaper]{IEEEtran}
\usepackage[cmex10]{amsmath}
\usepackage{amssymb,amsfonts,textcomp}
\usepackage{mathtools}
\usepackage{latexsym}
\usepackage{graphicx}
\usepackage{subfigure}
\usepackage{url}
\usepackage[plain]{algorithm}
\usepackage{paralist}

\usepackage[nospace, compress]{cite}

\newtheorem{example}{Example}

\begin{document}

\title{Exploring heterogeneity of unreliable machines for p2p backup}

\author{
\IEEEauthorblockN{Piotr
   Skowron (corresponding author)
}
\IEEEauthorblockA{Faculty of Mathematics, Informatics and Mechanics\\
  University of Warsaw \\
  Email: p.skowron@mimuw.edu.pl
}
\and
\IEEEauthorblockN{Krzysztof
  Rzadca
}
\IEEEauthorblockA{Faculty of Mathematics, Informatics and Mechanics\\
  University of Warsaw \\
  Email: krzadca@mimuw.edu.pl
}

}

\maketitle

\begin{abstract}
P2P architecture is a viable option for enterprise backup. In contrast to dedicated backup servers, nowadays a standard solution, making backups directly on organization's workstations should be cheaper (as existing hardware is used), more efficient (as there is no single bottleneck server) and more reliable (as the machines are geographically dispersed).

We present the architecture of a p2p backup system that uses pairwise replication contracts between a data owner and a replicator. In contrast to standard p2p storage systems using directly a DHT, the contracts allow our system to optimize replicas' placement depending on a specific optimization strategy,
and so to take advantage of the heterogeneity of the machines and the network. Such optimization is particularly appealing in the context of backup: replicas can be geographically dispersed, the load sent over the network can be minimized, or the optimization goal can be to minimize the backup/restore time.
However, managing the contracts, keeping them consistent and adjusting them in response to dynamically changing environment is challenging.

We built a scientific prototype and ran the experiments on 150 workstations in the university's computer laboratories and, separately, on 50 PlanetLab nodes.
We found out that the main factor affecting the quality of the system is the availability of the machines.
Yet, our main conclusion is that it is possible to build an efficient and reliable backup system on highly unreliable machines (our computers had just 13\% average availability).


\end{abstract}

\section{Introduction}\label{sec::introduction}
Large corporations, medium and small enterprises, universities, research centers and common computer users are all interested in protecting their data against hardware failures.
The most common approach to protecting data is to keep the backup copies on tape drives, specially designated storage systems, or to buy cloud storage space. All such solutions are highly reliable, but also expensive. In 2013 the costs of renting 1TB of cloud storage per year from Amazon, Google, Rackspace or Dropbox is approximately \$1000. Additionally, for some organizations, internal data handling policies require that data cannot be stored externally. The price of a single backup server with raw capacity of 14TB often exceeds \$12,000. A tape-based backup system for 14TB costs about \$7,000. This figures do not include additional costs of service, maintenance and energy.
With a large number of workstations that must be replicated at a single server, the server may become a bottleneck and may be not able to offer satisfactory throughput; also, performance can be degraded by network congestion. More scalable solutions exist, but are even more expensive.
Yet, the market for the backup solutions is vast. DataDomain, a company providing the modern backup systems, had in 2009 over 3.000 customers and over 8.000 systems deployed~\cite{dataDomainCustomers}. In the same year the company was bought by EMC for \$2,4 billion.

There is still a need for cheaper alternatives for enterprise backup. On one hand, a significant research effort focuses on the optimization techniques for dedicated backup servers, such as 
deduplication techniques~\cite{Meyer:2011:SPD:1960475.1960476, Dong:2011:TSD:1960475.1960477}; or erasure codes~\cite{conf/infocom/OggierD11, conf/icdcn/Pamies-JuarezOD13}. On the other hand, a p2p architecture can be explored in the context enterprise backup. Common PCs are cheaper than reliable servers. Also, in many cases, the unused disk space on the desktop workstations can be used without additional costs (Adya~et~al.~\cite{farsite} discovered a tendency that the unused disk space on the desktop workstations is growing every year; the Moore's law for hard disks capacities, first formulated by Kryder~\cite{citeulike:8870751} still holds). The bandwidth of the nodes connected in a distributed way scales much better than a single server; the load on the network is more evenly distributed causing less bottlenecks. The system can take advantage of the geographical dispersion of the resources, thus offering better protection in case of natural disasters (e.g., fire, flood) or theft. Finally, p2p solutions have already proved to work well in the enterprise environment (GFS~\cite{googlefs}, MapReduce~\cite{Dean:2008:MSD:1327452.1327492}, Astrolabe~\cite{Renesse01astrolabe:a}, DHT~\cite{dHash} used in HYDRAstor~\cite{hydraStore}, etc.).

Indeed, many p2p storage systems have been already built~\cite{ivyP2PFileSystem, pastisP2PFileSystem, igorFs, bitVault, bigTable, oceanStore, farsiteRetrospection, friendstore, freenet, wualaStudy, pangea}. The deduplication techniques get adapted for p2p storage systems~\cite{XingLD10, conf/p2p/PapapetrouRSN10}. There are new erasure codes more suitable for p2p systems~\cite{conf/infocom/OggierD11}.
Finally, there are many theoretical models for data placement optimizing data availability~\cite{Rzadca2010ReplicaPlacementin, conf/p2p/Pamies-JuarezLA11, Bernard2009OptimizingPeer-to-PeerBackup, douceur2001competitive, chun2006efficient, Bhagwan2002Replicationstrategieshighly,rodrigues2005high} and backup/restore performance~\cite{p2pDataTransferScheduling, conf/p2p/Pamies-JuarezLA10} in p2p storage systems.
However, real systems do not fully take advantage of the p2p architecture. There is a gap between theoretical models and real implementations.
There are systems (e.g., OceanStore~\cite{oceanStore} and Cleversafe) that distribute data between geographically remote servers. These systems could be used for backup, but they both use dedicated servers, which stays in the contrast with our primary goal of creating cheap backup system based on existing, unreliable machines.

There are p2p storage systems designed to work on unreliable machines; perhaps the most known such a system is Farsite~\cite{farsiteRetrospection} -- a 6-years long Microsoft's project. However, Farsite offers much more than a simple backup. As a complete distributed file system, Farsite must deal with parallel accesses to data, must manage the file system namespace, and ensure that frequently accessed data is highly available. Such requirements force additional complexity and many architectural limitations that do not exist in case of  backup system. On the other hand, since data backup is not a primary use-case, Farsite does not focus on implementing replica placement strategies (e.g. geographical dispersion of replicas, ensuring that data are backed up within a given time window etc.). For more discussion on the p2p storage systems we refer the reader to the next section.

Bridging the gap between many theoretical models~\cite{p2pDataTransferScheduling, conf/p2p/Pamies-JuarezLA10, Bernard2009OptimizingPeer-to-PeerBackup, douceur2001competitive, chun2006efficient, Bhagwan2002Replicationstrategieshighly,rodrigues2005high, Rzadca2010ReplicaPlacementin, conf/p2p/Pamies-JuarezLA11} and prototype implementations, we asked the following question: Is it possible to implement various data placement strategies especially when machines are unreliable? Certainly, there are more challenges than in the case of centralized or highly-available systems. The machines' unreliability, and perhaps low availability, requires  data locations to change dynamically. Is it difficult to continuously optimize the data placement with such assumptions? And, finally, is it difficult to take advantage of the machines nad network heterogeneity?

\textbf{Our main contribution is the following:} (i) We present an architecture of a prototype storage system that uses pairwise (bilateral) replication contracts for storing data. (ii) We show that we can efficiently manage the contracts and ensure efficient backup
even under significant peers' unavailability. Our scientific prototype is evaluated in a real distributed environment.

We built a scientific prototype that replicates user data on different workstations of the organization. In our prototype, the machines that enter the system besides the standard activities also keep replicas of data of other peers. We assume that the workstations are heterogeneous and prone to failures, in particular: (i) hardware might be heterogeneous and inefficient; (ii) the workstations may have variable amount of unused disk space (the space that is available for keeping replicas); (iii) the workstations are not always available -- computers might stay powered on, or be powered off, when not used by anyone (transient failures) (iv) they may experience permanent failures after which it is not possible to recover data stored on a machine.

In contrast with fixed data placement (storing data in a DHT \cite{ivyP2PFileSystem, pastisP2PFileSystem, igorFs, bitVault, bigTable}), our replication is based on the storage contracts between an owner of the data and its replicators. A contract for storing a data chunk of the owner $i$ on the replicator $j$ is a promise made by $j$ to keep $i$'s data chunk for a certain amount of time. Until the contract expires, it cannot be dropped by $j$; but it can be revoked by $i$. Since every data chunk is associated with a list of storage contract, each chunk can be placed at any location (the location depends on the placement strategy). This contract-based architecture can be exploited in two ways. First, the contracts form an unstructured, decentralized architecture that enables to optimize replica placement, making the system both more robust and able to take advantage of network and hardware configuration.
Second, contracts also allow strategies for replica placement that are incentive-compatible, such as mutual storage contracts~\cite{Rzadca2010ReplicaPlacementin, cox2003samsara}. To the best of our knowledge, all previous literature on mutual contracts focuses on theoretical analysis only. We complement these theoretical works, by presenting an architecture of a contract-based storage system. Yet, in this paper, for the sake of concreteness, we focus on optimization of replica placement for p2p backup in a single organization, where incentives are not needed.

We have implemented a prototype. The prototype (with the source code) is available for download with an open-source lincense at \url{http://www.mimuw.edu.pl/~krzadca/nebulostore/software.html}. 
We tested our prototype on 150 computers in students' computer laboratories; and on 50 machines in Planet-Lab. The lab environment might be considered as a worst-case scenario for an enterprise network, as the computers have just 13\% average availability and are frequently rebooted. Moreover, we assumed that all the local data is modified daily. 

\textbf{The results of our work show that:} (i) in a p2p backup system we are able to efficiently transfer data chunks -- the bandwidth of such a system scales linearly with the number of machines. (ii) Even on machines with very low availability we are able to efficiently optimize placement of the replicas. We verified two different placement strategies (where the optimization goal was either to finish the backup of each data chunk within required backup window, or to enforce a certain geographical disperison of the replicas) in two different settings. This leads to our main conclusion: (iii) It is possible to create an efficient p2p backup system and to take advantage from the peer's heterogeneity. 
(iv) Hardware unavailability has a significant impact on the performance of the backup; because of unavailability the time needed for direct communication of two peers can be long (on average 20h). We call this effect the \emph{cost of unavailability}.
Our measurements confirm the simulation results of Sharma~et~al.~\cite{friendstoreAnalysis} and Tinedo~et~al.~\cite{conf/p2p/TinedoAL12}.

Since our results are supported not only by the simulations, but also by measurements of a implementation on a real system, we consider them as the proof of the concept that an efficient p2p backup systems can be created and that the heterogeneity of the machines in such a system can be explored.

\section{Related Work}\label{sec::related-work}

In this section, we review the related commercial projects and scientific approaches to data replication in distributed systems.

HYDRAstor~\cite{hydraStore} and Data Domain~\cite{dataDomain} are commercial distributed storage systems, which use data deduplication to increase amount of the virtual disk space.

Many papers analyze various aspects of p2p storage by either simulation or mathematical modeling. 
Usually, the analysis focuses on probabilistic analysis of data availability in presence of peers' failures (e.g., \cite{Bernard2009OptimizingPeer-to-PeerBackup}). 
Douceur~et~al.\cite{douceur2001competitive}, similarly to our system, optimize availability of a set of files over a pool of hosts with given availability: theoretical as well as simulation results are provided for file availability.
Chun~et~al.~\cite{chun2006efficient} studies by simulation durability and availability in a large scale storage system. 
Bhagwan~et~al.~\cite{Bhagwan2002Replicationstrategieshighly} and Rodrigues and Liskov~\cite{rodrigues2005high} show basic analytical models and simulation results for data availability under replication and erasure coding. 
Finding the schedule of the transfers which minimizes the restore time and analysis of the impact of the size of the set of the replicators on the restore time is described by Toka~et~al.~\cite{p2pDataTransferScheduling}. Pamies-Juarez~et~al.~\cite{conf/p2p/Pamies-JuarezLA10} studies the impact of the redundancy on the data retrieval time.
Our paper complements these works by, firstly, presenting the architecture that allows for implementing placement strategies; secondly, by considering other measures of efficiency; and, thirdly, by proving that various optimization strategies can be accomplished in unreliable environment.

As the context of this work is data backup in a single organization, we do not analyze incentives to participate in the system. 
However, to store the data, our system relies on agreements (contracts) between peers.  
In contrast, in DHT-based storage systems contracts are (implicitly) made between a peer and the system as a whole.  
Thus, our architecture naturally supports methods of organization that emphasize incentives for high availability, such as mutual storage contracts~\cite{Rzadca2010ReplicaPlacementin, cox2003samsara} (also these using asymmetric contracts~\cite{conf/p2p/Pamies-JuarezLA11}).
It is worth mentioning that some papers explore the social interconnections while choosing the replica locations~\cite{friendstore}; the tradeoffs  between the redundancy, data availability and the ability to place data on the trusted nodes is analyzed  by Sharma~et~al.~\cite{friendstoreAnalysis} and Tinedo~et~al.~\cite{conf/p2p/TinedoAL12}. These methods can also be adopted for our system.

Many p2p file systems~\cite{ivyP2PFileSystem, pastisP2PFileSystem, igorFs, bitVault, bigTable}, use storage and routing based on a DHT~\cite{dHash, pastryDHT}. The address of the block, which is a hash of its content, fully determines the locations of its replicas. Thus, such architecture is less suitable for balancing the load on replicating workstations, or for optimizing placement of replicas. While these solutions focus on consistency of the data being modified by multiple users, this paper focuses on the issue of the best replication of the data, which cannot be modified by anyone apart from its owner.

OceanStore~\cite{oceanStore} and Cleversafe
propose the idea of spreading replicas into geographically remote locations achieving the effect of a deep archival storage. These systems combine software solutions with a specially designed infrastructure that consists of numerous, geographically-distributed servers. The main contribution of these systems, from our perspective, is the resignation from a common DHT and the introduction of a new assumption that any piece of data can be possibly located at any server. These systems, however, do not discuss the issue of replicating data on the ordinary workstations (which are, in contrast to the servers, frequently leaving and joining the network) and do not present any means allowing to handle such dynamism.

Wuala~\cite{wualaStudy}
moved one step further by proposing a distributed storage based not only on a specially dedicated infrastructure, but also including a cloud of workstations of users who install Wuala application. However, since late 2011, Wuala no longer supports p2p storage. The idea of using a hybrid architecture of central servers and user machines, called in the context of backup as peer-assisted backup is also explored by Toka~et~al.~\cite{peerAssistedApproach}.
Other p2p backup software include Backup P2P~\footnote{sourceforge.net/projects/p2pbackupsmile/}, Zoogmo~\footnote{zoogmo.wordpress.com}, or ColonyFs~\footnote{launchpad.net/colonyfs}.

FreeNet~\cite{freenet}
is a p2p application that exposes the interface of a file system. Its main design requirement is to ensure anonymity of both authors and readers. The underlying protocol relies on proximity-based caching. When a data item is no longer used, it can be removed from a caching location. Similarly, in Pangea \cite{pangea}  a replica  is created whenever and wherever a data is accessed.

Farsite~\cite{farsiteRetrospection} was a Microsoft's 6-years long project aimed at creating distributed file system for sharing data between thousands of users. The retrospective from the project~\cite{farsiteRetrospection} gave us the feeling of following a good direction. Firstly, the authors emphasize that real scalability must face the problem of constant failures in the network. Secondly, they claim that in a scalable system, manual administration must not increase with the size of the network; we followed the both requirements when formulating our hypothesis.

There are a few substantial differences between Farsite and our prototype implementation. Most importantly, Farsite's architecture does not rely on mutual contracts, which allow us to implement both incentives and mechanisms ignoring the black listed peers.

In Farsite updates of data are committed locally and the changes are appended to the log (similarly to Coda \cite{codaDisconnectedOps}). The log is sent to a group of peers responsible for managing a subset of a global name space (called directory group), which periodically broadcast log to the all group members. As directory group uses Byzantine Fault Tolerant protocol \cite{farsiteByzantine}
no file can be modified if one third or more of the group members is faulty. Since we consider a backup system in which data is modified only by the owner
we are able to gain in flexibility and robustness. In our asynchronous updates mechanism, every peer has associated group of synchro-peers managing its asynchronous messages. Synchro-peers are independent of replicas, which results in a desired property that every peer can keep replicas for any chunk of data. Thus, replicas can be chosen so that they constitute the most profitable replication group.

Farsite is a distributed file system and many of its use cases cause the greater complexity of the system. On the other hand, as Farsite is not a backup system, it does not support backup-specific requirements like placing replicas in geographically distributed locations, optimization of the backup/restore time, etc.


\section{System Architecture}\label{sec::architecture}
Our system uses a mixed architecture that stores control information, meta-data and data in three different ways.
The control information that allows peers to locate and to connect to each other 
must be located efficiently --- hence we use a DHT as a storage mechanism. 
In contrast, each peer is responsible for finding and managing peers who replicate its data (its \emph{replicators}).
Such replication contracts enable us to optimize replica placement and thus to tune replication to a specific network configuration.
The meta-data describing replication contracts are kept by both the data owner and the replicator.
Chunks of data are kept in an unstructured overlay; concrete locations are described by the meta-data.



\subsection{Control information}\label{sec::metadata}
The basic attributes of a peer are kept in a structure called \textit{PeerDescriptor}. 
For each peer, its PeerDescriptor contains:

\begin{list}{$\bullet$}{\setlength{\leftmargin}{8pt} \setlength{\labelwidth}{0pt}}
        \setlength{\itemsep}{2pt}

\item identification information (public key); 

\item information needed to connect to this peer (a current IP address and a port of an instance of our software running on a workstation) and user account name in the operating system (account name is required by the current implementation of data transmission layer --- see Section \ref{sec::dataTransmission});

\item identifiers of \emph{synchro-peers} (see Section~\ref{sec::asynchronous-mess}). 
\end{list}

PeerDescriptors of all peers are kept in a highly replicated DHT: peer's ID (a hash of its public key) is hashed to its PeerDescriptor. 
As the size of the control information is small, we are able to afford strong replication (compared to a generic DHT). 
Thus, instead of a single peer, a number of peers is responsible for keeping data hashed to a part of the key-space.

\subsection{Replication contracts}\label{sec::catalog}
The main goal of our prototype is to support nontrivial replica placement strategies; we need to be able to store any replica at any peer. 
This architecture contrasts with content addressable storage systems that put each chunk of data under an address that is fully determined by the chunk's unique identifier (e.g. hash of its content).
As a trade-off for flexibility of placing replicas at any location, we need a mechanism to locate data.

In our system each peer keeps information about replica placement of its data chunks in an index structure called \textit{DataCatalog}.  
For each data chunk, the catalog stores information about:
\begin{inparaenum}[(i)]
\item identifiers of the peers that keep replicas of the data chunk (hereinafter \textit{chunk replicators});
\item size of the chunk;
\item version number of the chunk.
\end{inparaenum}

Additionally, each peer keeps information about data chunks it replicates. 
As each storage contract is kept in exactly two places (the owner and the replicator), contracts are consistent and it is easy to retrieve lost metadata (the DataCatalog). 
Because peers are unreliable, the process of contracts negotiation can break at any point, possibly leading to two types of inconsistency: an owner $o$ believes $j$ is its replicator, while $j$ is not aware of such a contract; or a peer $j$ believes to be $o$'s replicator, while $o$ is not aware of such a contract. Contracts negotiation is however idempotent and because the contracts are kept both by owner and replicator, such inconsistencies can be easily fixed. Each peer periodically sends messages to its replicators with the believed contracts (and versions of data chunks, which allows the replicators to update the chunks of data which are out of date). Each replicator periodically sends similar information to the appropriate owners. Detected inconsistencies can be resolved either by adopting the owner's state; or by always accepting a replication agreement.

The DataCatalog is persisted in a file but it is not replicated between peers. In this way we avoid the additional overhead of updating the catalog at the remote locations, when the contract for any chunk is changed; because machines are unreliable, in many cases we even would not be able to update DataCatalog at the remote peers as they can be simply unavailable. On the other hand, both owners and replicators are aware of all their storage contracts. When the local data of any peer is lost, the peer (the owner) gossips the information about the failure.  The replicators answer the gossip message with the information about the contracts; the owner uses this information to rebuild DataCatalog. Once DataCatalog is reconstructed, the owner locates and rebuilds all missing data. Since the DataCatalog is persistent, its reconstruction is required only in case of non-transient failure; thus it does not cause much overhead.

Alternatively, in an enterprise environment where a (replicated) server is an affordable option, the meta-data can be kept centrally (in primary memory for faster access). This solution is, however, less scalable.

We designed the mechanism that is responsible for relocating or additionally replicating data chunks that are weakly replicated according to the given abstract metric. The specific metric used in our evaluation takes into account peers' availability, bandwidth and geographic distribution; it tries to keep all but one replicas as close as possible not to overload the network and to keep one replica in remote location for additional safety (for location-dependent failures, such as fire, flood, etc.) . The metric also balances the load on the machines so that each data chunk can be replicated within the required \emph{backup window} (the time requirement for each chunk to be backed up). The precise metric is described in Section~\ref{sec::replica-placement}. The optimization mechanism is based on hill-climbing --- in consecutive steps, each peer performs locally optimal changes of the contracts.
The optimization can proceed even if large fraction of peers is unavailable; to perform a single local optimization we require only 3 peers to be available. Thus the mechanism is suitable for unreliable environments.

When nodes parameters (e.g. availability) change, or when a large number of nodes is added to the system, the contracts are renegotiated.
If each such change resulted in data migration, the network and the hosts could easily become overloaded. Therefore the process of changing a contract is more elaborate.
The contracts are allowed to change frequently but such changes do not require data migration. Such temporary contracts are periodically (e.g., daily) committed; after a contract is committed, the data is migrated. The complete mechanism involves some additional details as it must take into account also possible communication failures -- see Section~\ref{sec::replica-placement}.

\subsection{Data Transmission and Updates}\label{sec::dataTransmission}
Every member of the network, before placing its replicas at a remote peer, must obtain this peer's permission. 
Once the peers reach an agreement, they mutually authorize each other using identities (public keys) available in peers' \emph{PeerDescriptor}s (stored in the DHT). 

The data is transmitted in an encrypted connection. 
In the current implementation, we use standard Linux tools for data transfers. Each peer runs a ssh daemon that acts as a server that accepts connections of data owners. When a peer initiates connection to transfer its data, it uses scp as a client.

When owner modifies its local copy, the updated chunk must be propagated to the network.
The replicators are informed of the changed versions of the data chunks through periodic control messages (the versions numbers are attached to the messages containing contracts sent between the owner and the replicator, described in the previous subsection). The unavailable peers are informed about the changes of data chunks through asynchronous messages, described below. Once the replicator finds there is a new version of the data chunk it replicates, it downloads the new version either from the owner or from the other replicators. Note that if the data owner were responsible for uploading the new version to the replicators, a successful transfer would require both the owner and the replicator to be available. In our solution, the replicator is responsible for keeping replicas up to date so we only require that the replicator and any other replicator or the owner is available. 

Unlike common backup systems, our system stores only the last version of each data chunk.
A system storing many previous versions may be built in the same way as e.g. version control software (svn, git) uses a standard filesystem; more specifically, the previous versions (or the deltas) can be kept in the same data chunk; or the deltas can be kept in separate data chunks.

\subsection{Asynchronous/Off-line Messaging}\label{sec::asynchronous-mess}

We assume that the workstations may be unavailable for some time just because they are temporarily powered off. In contrast to many distributed storage systems (e.g., GFS \cite{googlefs}), in such a case our system does not rebuild the missing replica immediately, in order not to generate unnecessarily load on other machines nor the network. Instead, when the unavailable peer eventually joins back the network, it efficiently updates its replicas.
To inform the unavailable peers about the new version numbers of their replicas and about the contracts, we use asynchronous messaging.
The control messages sent to the peer that is currently unavailable are cached at, so called, \emph{synchro-peers}. We use the idea of group communication for synchronizing the messages within each (small) group of synchro-peers. As opposed to Defrance~et~al.~\cite{edgeBuffering}, who present the mechanism of caching the messages on routers, we chose to design the concept of synchro-peers to limit the costs of the additional hardware.

An asynchronous message from $i$ to $j$ is sent to the \emph{synchro-peers} of $j$. 
Synchro-peers is a set of peers, defined for every peer $j$ ($j$ is called in this context a \emph{target peer})
that keep asynchronous messages for $j$. Synchro-peers of $j$ include $j$, so every message will be delivered to the target peer by the same means as it is delivered to the other synchro-peers.
Each synchro-peer periodically tries to send the asynchronous message to the synchro-peers that have not yet received the message; the IDs of synchro-peers that have not yet received the message are attached to the message (thus, the same message can be delivered multiple times to the same peer).

The consecutive messages between any two peers are versioned with sequential numbers (logical clocks). If any synchro-peer $k$ has not managed to send a message $m_{1}(i \to j)$ to the all requested synchro-peers before receiving a subsequent message $m_{2}(i \to j)$ with a higher version number, then the synchro-peer drops $m_1(i \to j)$, as the new message already contains more version numbers of the data chunks. Thus, the expected number of messages that are waiting for delivery on a single synchro-peer is bounded by $|R(\cdot)| \cdot |S(\cdot)|$, where $|R(\cdot)|$ is the average number of replicators per peer and $|S(\cdot)|$ is the number of synchro-peers per peer. The group of synchro-peers is small (5 in our experiments), the messages contain only the version numbers of the data chunks (thus, the messages are small as well), and the old undelivered messages can be safely replaced with by the newer versions of the messages. As the result the mechanism of asynchronous messaging is cheap from the perspective of the system.

The target peer executes the commands from the message immediately after the first reception but it remembers for each sender the latest version number of the received message. This information protects against the multiple execution of the orders from a single message.

The mechanism of versioning through asynchronous messages reduce the amount of information replicators keep about the structure of replication contracts. In an alternative solution, replicators synchronize directly between each other. However, this requires replicators to know IDs of all other replicators for each data chunk they store. If this information is stored at replicators, changes in replication contracts require multiple updates; if it is stored as meta-data, the size of the meta-data becomes proportional to the number of data chunks in the system. Moreover, exchanging the messages between all replicators is highly inefficient. With asynchronous messages, the owner keeps the information about its replication contracts: updates are simple and meta-data is small.

Using asynchronous messaging has two advantages. First, the asynchronous message is delivered with high probability even when the sender is unavailable. Second, the data may be downloaded concurrently from multiple replicators.

\section{Replica placement}\label{sec::replica-placement}
The goal of a replica placement policy is to find and dynamically adapt the locations of the replicas in response to changing conditions (new peers joining, permanent failures, changing characteristics of existing replicators).  
Finding possible locations for replicas is not trivial, given that peers differ in availability and amounts of free disk space.
Moreover, the replica placement policy should take advantage of peers' heterogeneity (like availability, geographic locations, etc.). 

Our policy consists of two main parts.
First, a utility function (in short \emph{utility}) scores and compares replica placements. A utility is a function that, for a given data owner and a set of possible replicators returns a score proportional to expected quality of replicating data.
Second, a protocol manages replica placement in the network in order to maximize the utility of the currently worst placement ($\max \min$ optimization).

\subsection{Utility function}\label{sec::utility-function}

A user of a backup application is interested in the resiliency level of her data (defined by the desired number of replicas $N_{r}$ and their proper geographic distribution); 
and the time needed to retrieve the data in case the local copy is lost (expressed as the desired data read time $Des(T_{r})$). Additionally, a user must be able to backup her data (propagate the local updates to replicas) during the time the user is on-line (expressed as a backup window $Des(T_b)$).

Utility function $U: P_k \to \mathcal{R}$ is a scoring function mapping a replica placement $P_k = P(d_k) = (o(d_k), R(d_k))$ to its score $u_k = U(P_k)$.

The average duration of data backup to replicator $j$, $E(T_b, j)$ is estimated by:
\begin{align}\label{eq::backup-duration}
E(T_b, j) = \frac{\sum_{k:j \in R(d_k)}{size(d_k)}}{p_{av}(j)B_{j}}
\end{align}
Backup duration is proportional to the congestion on the receiving peer $\sum_{k:j \in R(d_k)}{size(d_k)}$; and inversely proportional to the bandwidth $B_j$ that peer $j$ dedicates for the background backup activities ($B_j$ is bounded by network and disk bandwidth, but can be further reduced by the user). We use a simplified model that does not explicitly consider the network congestion, but this issue is addressed in the next subsection. Moreover, successful write on $j$ is possible only when $j$ is available (hence $p_{av}(j)$).

Assuming that restore operation has no priority over the backup, the average data restore duration $E(T_r, j)$ is computed in the same way.

The geographical distribution of the data is approximated by TTL values. Most of the replicas should be near the owner to reduce the network usage. $close_{max}$ denotes the desired distance for the ``nearby'' replicas. However, to cope with geographically correlated disasters, one replica should be far: its distance should be between $remote_{min}$ and $remote_{max}$.

The utility is a sum of utilities expressing geographic distribution, backup time (performance) and the number of replicas (with the former two treated essentially as constraints):
\begin{align}\label{eq::utility}
U(P_k) =  U_{geo}(P_k) - L \cdot ||R(d_k)| - N_{r}| - M \cdot U_{perf}(P_k) \text{,}
\end{align}
where $M$ and $L$ are (large) scaling factors. $L$ penalizes for insufficient number of replica. $M$ penalizes for backups that cannot be finished within the time window. If the backup cannot be done on time, it means for some data we can give no resiliency guarantees and so, even very good geographic distribution properties are useless.

The backup time penalty $U_{perf}(P_k)$ is the sum over the utilities per replica location:
\begin{align*}
U_{perf}(P_k) = \sum_{j \in R(d_k)}U_{perf}(j)
\end{align*}
where $U_{perf}(j)$ penalizes for insufficient backup window on $j$-th replicator:
\begin{align*}
U_{perf}(j) = \;\; & \textnormal{min}(Des(T_b) - E(T_b, j), 0) \\
& + \textnormal{min}(Des(T_r) - E(T_r, j), 0)
\end{align*}

The geographic utility $U_{geo}(P_k)$ considers both ``near'' and ``far'' replicas:
\begin{align*}
U_{geo}(P_k) =& \;\textnormal{min}(0, dist_{TTL}(j_{max}) - remote_{min}) + \\
& \textnormal{min}(0, remote_{max} - dist_{TTL}(j_{max})) +\\
& \sum_{\mathclap{j \in P_k - \{j_{max}\}}}\textnormal{min}(0, close_{max} - dist_{TTL}(j))
\end{align*}
where $j_{max}$ denotes the most distant location and $dist_{TTL}(j)$ denotes the TTL distance between $j$ and the data owner. 

Additionally, in order to limit data movement when $U(P_k)$ differs from $U(P_k')$ only by small value (in our experiments, 10\%) we treat these values as equal.

In an enterprise backup system, we assume that all data chunks are equally valuable. Thus, the utility of the whole system is the utility of the worst placement ($\max \min_k u_k$).



\subsection{Distributed optimization protocol}

We considered several approaches for maximizing system utility $\max \min_k u_k$. Probably the most straightforward idea is that each peer is responsible for its own utility $u_k$. This approach allows to implement game-theoretic strategies~\cite{Rzadca2010ReplicaPlacementin, cox2003samsara} that protect each peers' selfish interests. Game-theoretic strategies would give the system extra protection against malicious spammers. However, they have the following drawbacks: (i) every peer has to compete with the other participants; in particular, peers with low availability or bandwidth could never achieve satisfactory replication; (ii) even if contracts for these  peers are accepted at the cost of rejecting the contracts of the high utility peers, such frequent contracts rejections will result in protocol inefficiencies. Considering these drawbacks we decided to turn to a proactive approach described below.

Every peer $i$ with free storage space periodically chooses a data chunk $d_k$ with low utility, and proposes a new replication agreement with the data chunk's owner $o$ (peers share information on low utility data chunks in a distributed priority queue).
The owner either tentatively adds $i$ to its replication set $R(d_k)$ (if the number of replicas $|R(d_k)|$ is lower than the desired resiliency level $N_r$); or tentatively replaces $j \in R(d_k)$, one of its current replicas, with $i$ (all possible $j \in R(d_k)$ are tested). If the resulting utility $U(P_k')$ is higher than the current value $U(P_k)$, the owner $o$ tries to change the contracts (see the next section). When the owner rejects the proposition or when it is unavailable, $i$ puts $o$ in a (temporary)  taboo list in order to avoid bothering it later with the same proposition. 

As the result of continuous corrections of the replicas placements each peer can end up having replicas of different chunks at different peers. Such a machine has many replicators and their monitoring becomes expensive. However, the monitored information (the availability and the size of replicated data) are gossiped; thus the cost of distribution of information is independent on the number of replicators. On the other hand, storage contracts with multiple peers allow to parallelize data transfers and the cost of replicas rebuilding is amortized.

If every peer proposed storage for the owner of the data with the lowest utility, the owner would get overloaded with storage offers (and the remaining data chunks would be ignored).
Therefore, each peer sends a message to $o$ with probability $p_{p}$ such that $p_{p} = \frac{T}{N} \alpha$, where $N$ is the estimated number of peers in the network, $T$ is the duration of the period, mentioned before, and $\alpha$ is desired number of messages that peer wants to get in a time unit without being overloaded. 
Given such probability, the expected value of the number of messages, $E_{m}$, the owner of data gets in a time unit is: $E_{m} = p_{p} \cdot N \cdot \frac{1}{T} = \alpha$.

The system keeps the data chunks with the lowest utility in a distributed priority queue. 
In our prototype, we implemented the distributed priority queue by a gossip-based protocol. Each peer keeps a fixed number of data pieces with the lowest priorities. It updates this information with its own data pieces and distributes the information to the randomly chosen peers.

\subsection{Changing replication contracts}

The contracts in our system are continuously renegotiated. Each such change cannot result in in data migration not to overloaded the hosts nor the network. The efficient changes of the contracts are described below.

\subsubsection{Finding the best replicators}

Below, we describe two aspects of the protocol: revocation of inefficient contracts; and recovery from transient failures.

When peer $i$ offers its storage to data owner $o$, and if $o$ decides that $i$ should replace one of its existing replicators $j$ (as the resulting value of utility function $U$ is higher), then $o$ has to explicitly revoke the contract with $j$. Thus, changing location of the data of $o$, from $i$ to $j$, requires these three peers being on-line. The example below illustrates why revocation of the contracts cannot be realized asynchronously. 

\begin{example}
Consider peer $j$ storing many data chunks of several owners. From the perspective of each owner, as $j$ is comparably overloaded, any new peer joining the network is a better replicator than $j$. If the contracts could be revoked asynchronously, all the peers would revoke the contract on $j$ during its unavailability. Now $j$, having no data, can take over all data stored at some other peer $k$ during $k$'s unavailability, by offering storage space to the all data owners replicating their data at $k$. Such situation can repeat indefinitely. Each peer is not aware that $j$ has already revoked some of its contracts and that it is not overloaded any more.
\end{example}

However, if the existing replicator $j$ has low availability, on-line revocation of its contract is also improbable. Thus, an owner can revoke a contract with such a plow-available replica (e.g., the first decile of the population) also through an asynchronous message.

Additionally, because the peers are unreliable, the process of contracts negotiation can break at any point leading to inconsistency of the contracts. Two types of inconsistency are possible: an owner $o$ believes $j$ is its replicator, while $j$ is not aware of such contract; or a peer $j$ believes to be $o$'s replicator, while $o$ thinks $j$ is not. Contracts negotiation is however idempotent and inconsistent contracts can be easily fixed. Each peer periodically sends messages to its replicators with the believed contracts (and versions of data chunks, which allows the replicators to update the chunks of data which are out of date). Each replicator periodically sends similar information to the appropriate owners. Detected inconsistencies can be resolved either by adopting the owner's state; or by accepting a replication agreement.

\subsubsection{Committing contracts and transferring data}

In order to reduce the load on the network, replicators cannot change too often; but to maintain high performance, replicators must eventually follow the negotiated contracts. A \emph{non-committed} contract between an owner and a replicator is negotiated, but no data has been transferred. Contracts are committed periodically. For each data chunk, if there is a new contract (negotiated, but not committed), the contract is committed when the time that passed since the last committed contract for this chunk is large enough (e.g., 24 hours). This guarantees that the data is transferred at most once in each time period (e.g. at most once each 24 hours); but even when (non-committed) contracts change often, data is replicated (as the committed contracts represent a snapshot of utility optimization).

After committing a contract, the owner sends a message to the new replicator that requests data transfer. As soon as the new replicator downloads requested chunk, it sends an acknowledgment to the owner. Finally, the owner notifies the old location to remove the chunk.

\section{Experimental Evaluation of the Prototype}\label{sec::experiments}

\subsection{Experimental environment}

We performed the experiments in two environments: (i) computers in the faculty's student computer labs; and (ii) PlanetLab. We run the prototype software for over 4 weeks in the labs and 3 weeks in PlanetLab. Each computer acted as a full peer: owned some data and acted as a replicator. The data was considered as modified at the beginning of each day; thus each day we expected the system to perform a complete backup. If the transfer of a particular data chunk did not succeed within a day, the following day we transferred a newer version of the chunk. We used chunks of equal size -- 50MB.

The computers were centrally monitored; the central monitoring server experienced several failures which slightly influence the presented results (the real backup times are  slightly shorter than presented).

\subsubsection{Students computer lab}

We run our prototype software on all 150 machines from the students computer lab. The availability pattern might be considered as a worst case scenario for an enterprise setting. 
The lab is open from Mondays to Fridays between 8:30am and 8pm and on Saturdays between 9am and 2:30pm. The students frequently (i) switch off or (ii) reboot machines to start Windows; each day at 8pm the computers are (iii) switched off by the administrators (the machines are not automatically powered on the next day); each of these events was considered a transient failure. 

\begin{figure}[tb]
     \begin{center}
         \includegraphics[width=2.6in]{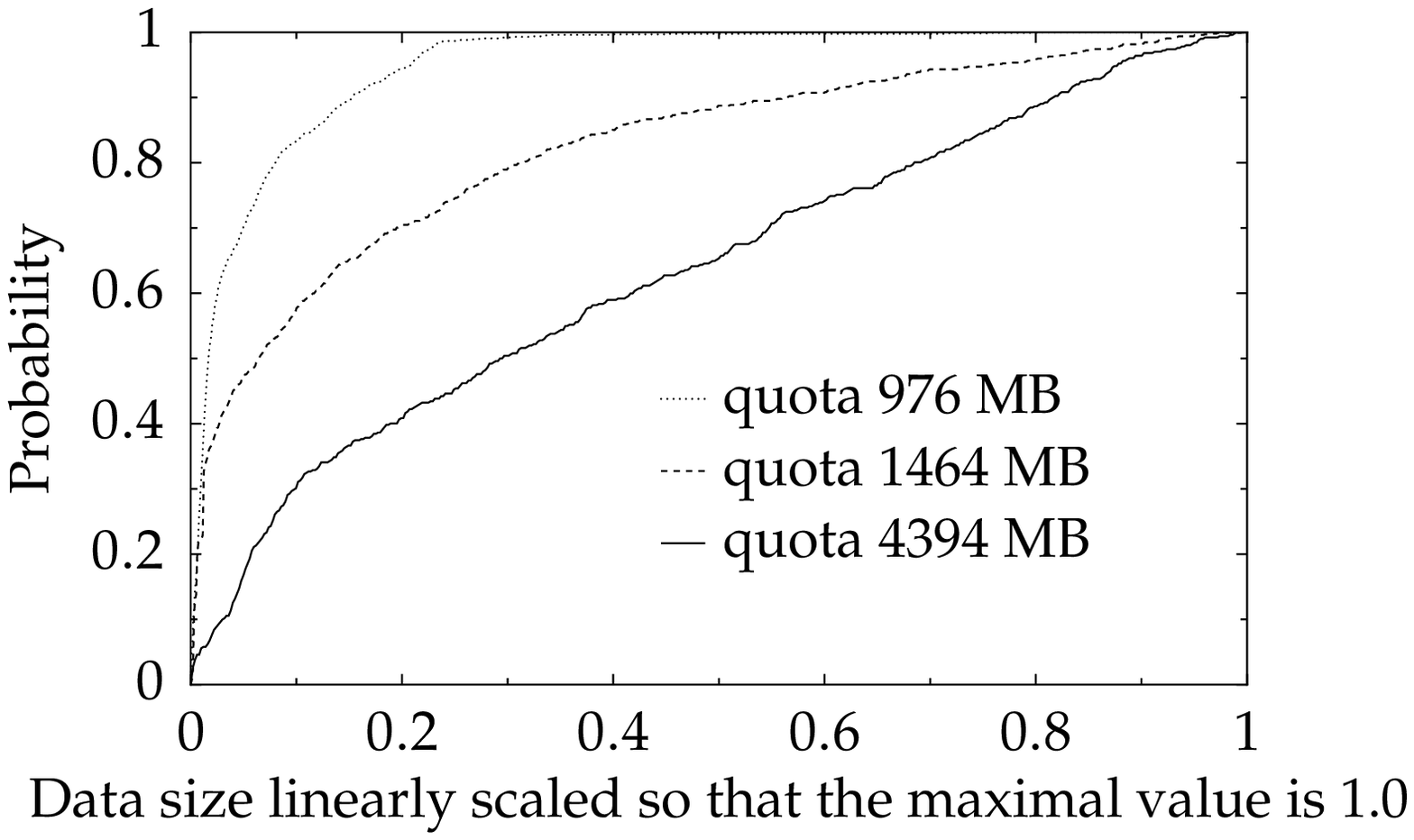}
     \end{center}
     \caption{The cumulative distribution functions of the sizes of students data for 3 groups of students, each having a specific quota value. The size of the data presented on abscissa is scaled so that the three distributions had the same maximal value equal to 1.}
     \label{fig:dataDistribution}
 \end{figure}

The amount of local data was sampled from the distribution of storage space used by the students on their home directories -- the students in our faculty are divided into 3 groups and each student is assigned an appropriate quota depended on her group affiliation. The distribution of data sizes for the three groups are presented in Figure \ref{fig:dataDistribution}.
We took the distribution of the sizes for the group with the highest quota and scaled this distribution so that the average value was 3GB. Thus the sizes of local data were varying approximately between 0 and 8GB.

The local storage space depended on machines' local hard disks; and varied between 10GB (50\% machines), 20GB (10\% machines), and 40GB (40\% machines).

\begin{figure*}[ht]
\begin{minipage}[b]{0.3\linewidth}
\centering
\includegraphics[width=\textwidth]{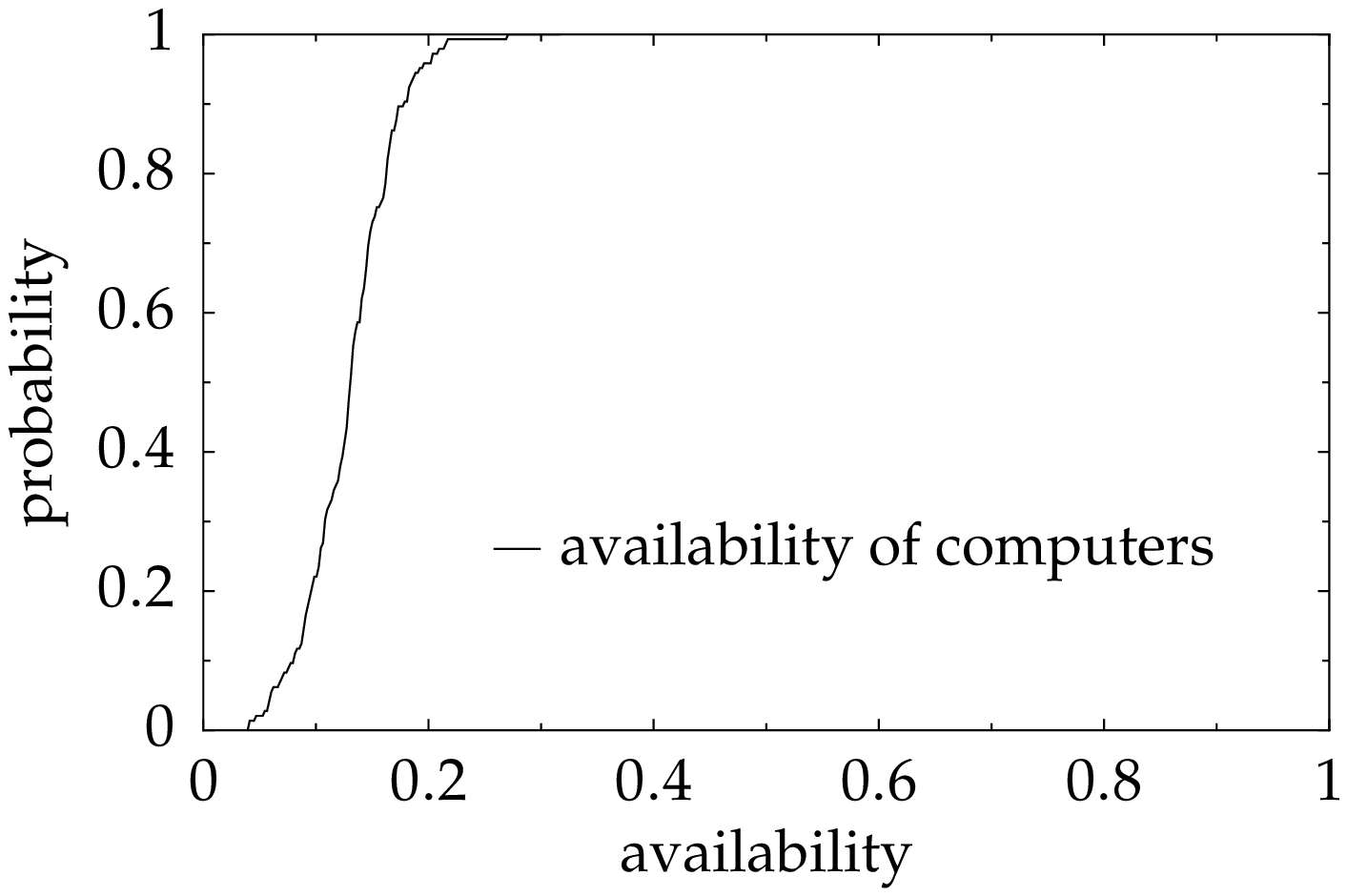}
\caption{The cumulative distribution function of the availability of the computers in students lab.}
\label{fig:labAvailability}
\end{minipage}
\hspace{0.5cm}
\begin{minipage}[b]{0.3\linewidth}
\centering
\includegraphics[width=\textwidth]{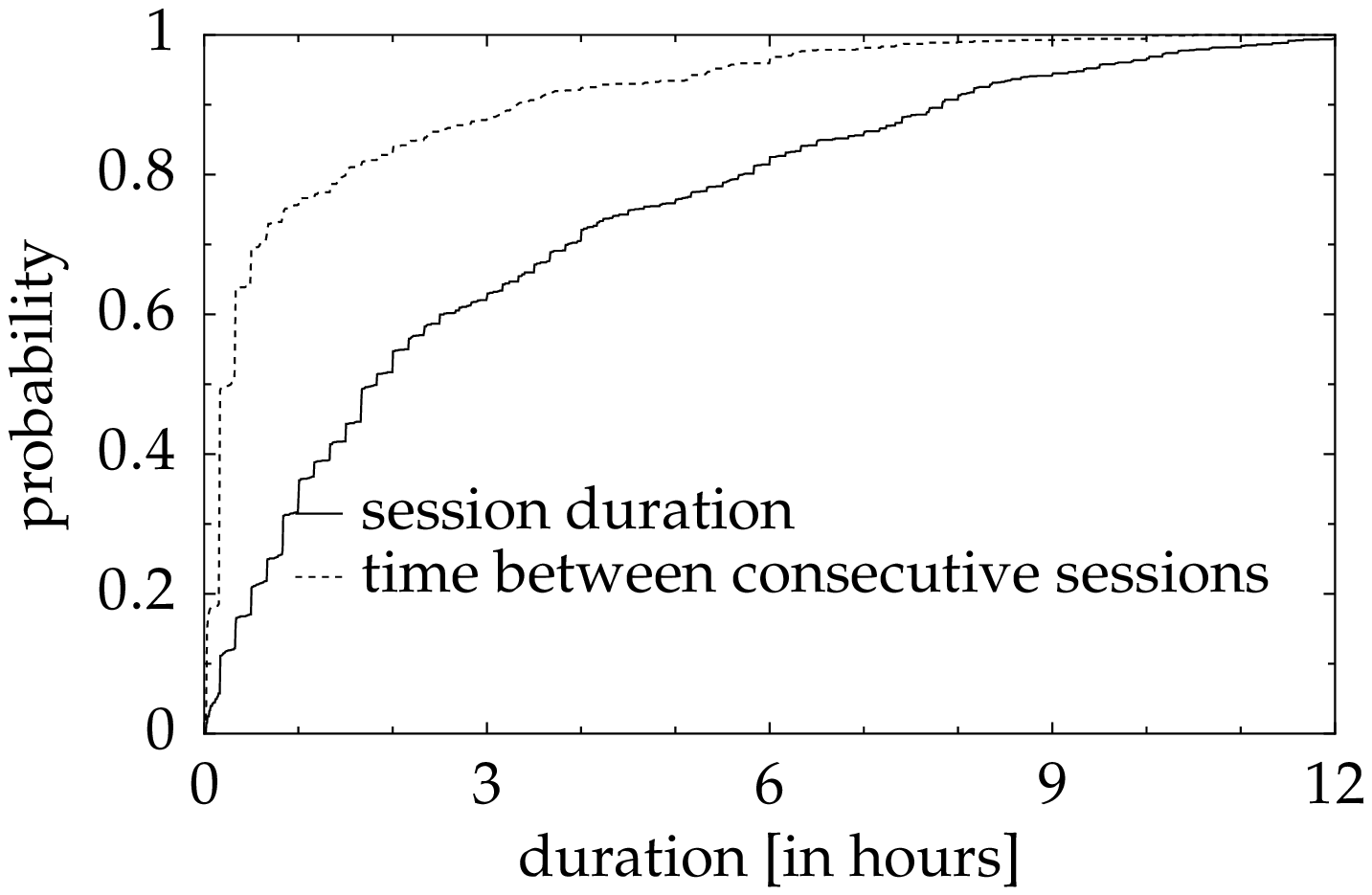}
\caption{The cumulative distribution function of the session durations and the times between consecutive sessions for the computers in students lab (the nights are filtered out).}
\label{fig:labUpTime}
\end{minipage}
\hspace{0.5cm}
\begin{minipage}[b]{0.3\linewidth}
\centering
\includegraphics[width=\textwidth]{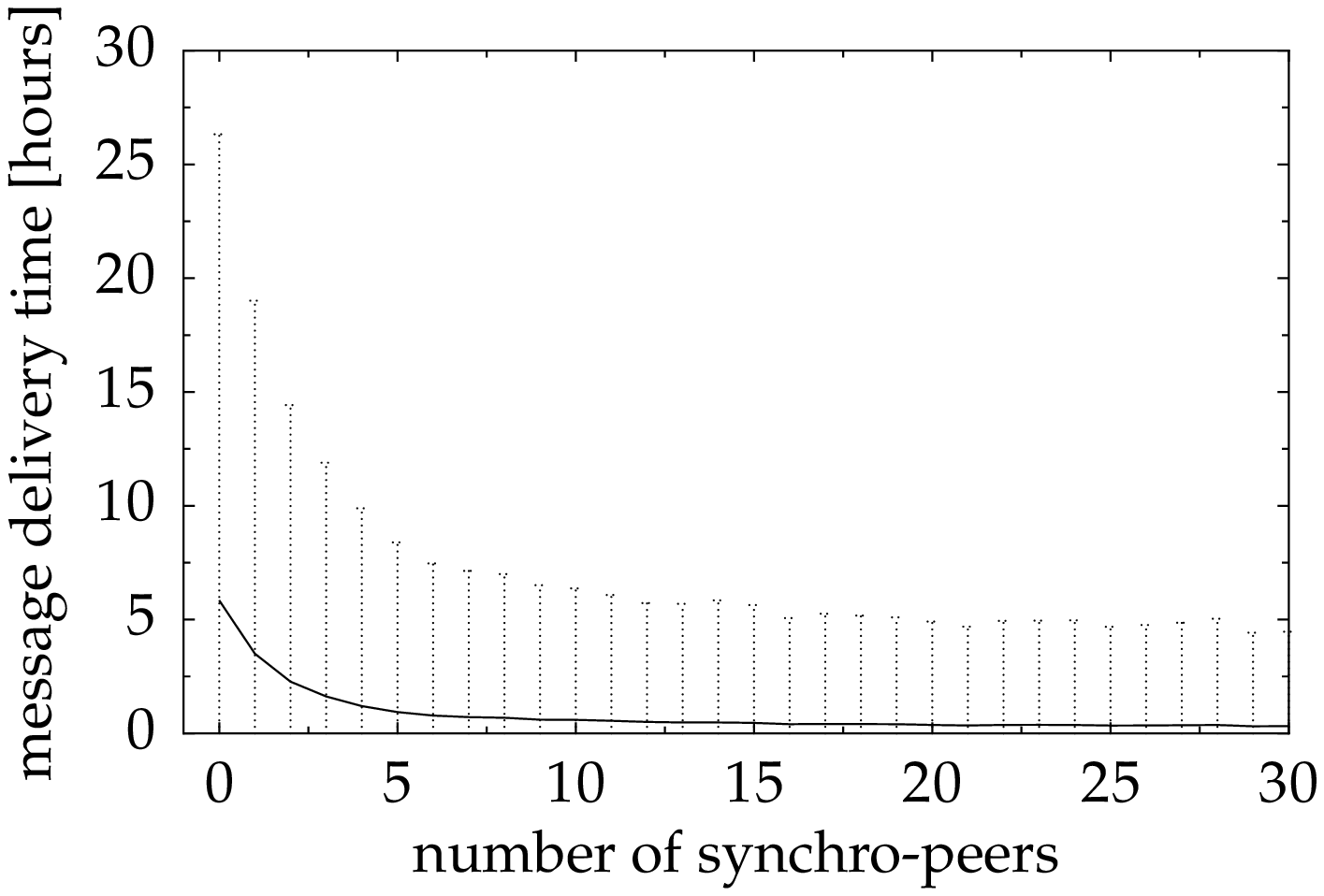}
\caption{The dependency between the number of synchro-peers and the delivery time of the asynchronous message. Delivery time measured from the first time of availability of the receiver after the message was sent.}
\label{fig:deliveryTime}
\end{minipage}

\begin{minipage}[b]{0.3\linewidth}
\centering
\includegraphics[width=\textwidth]{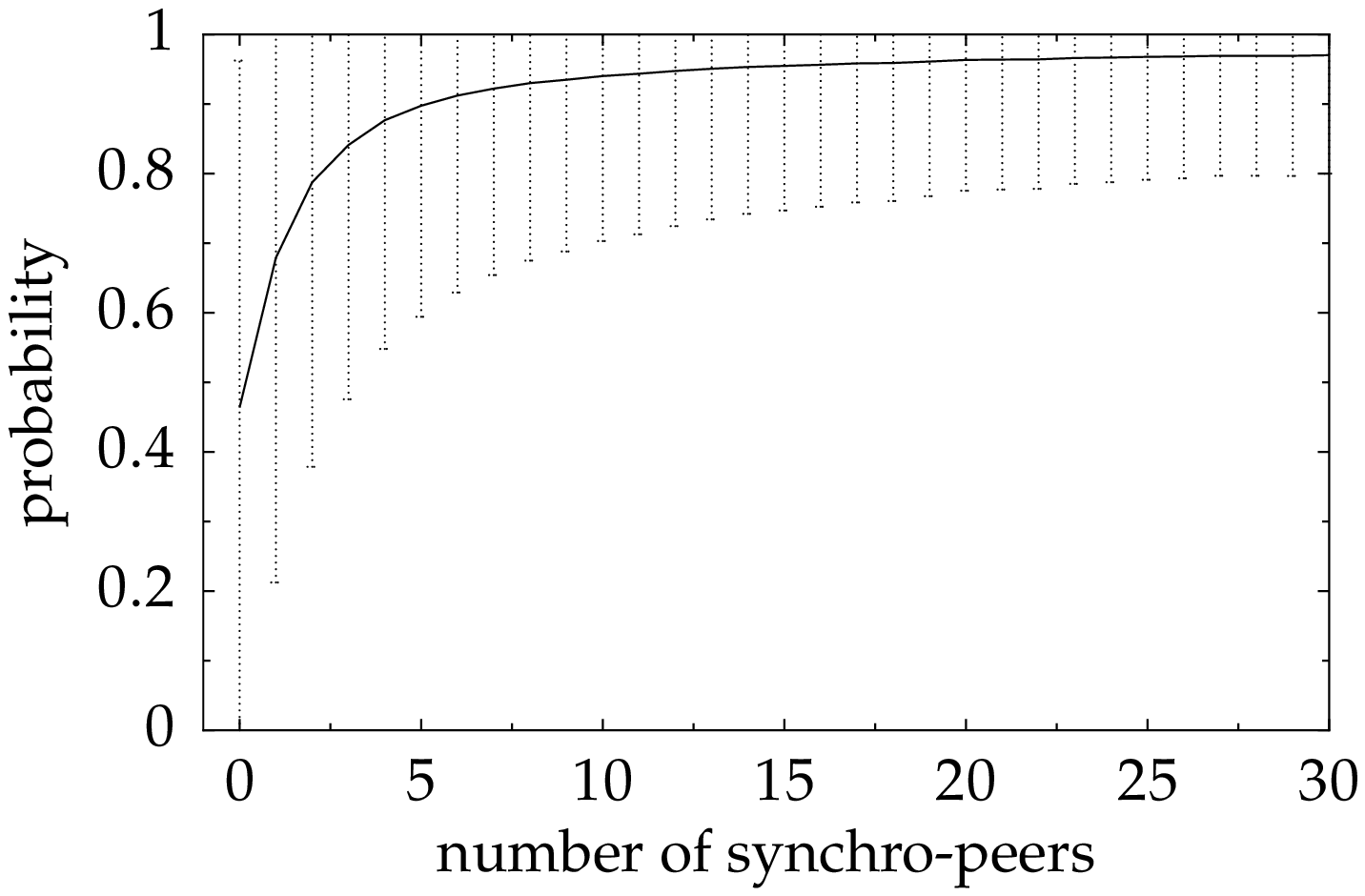}
\caption{The dependency between the number of synchro-peers and the probability of delivery a message to any synchro-peer.}
\label{fig:deliveryProbability}
\end{minipage}
\hspace{0.5cm}
\begin{minipage}[b]{0.3\linewidth}
\centering
\includegraphics[width=\textwidth]{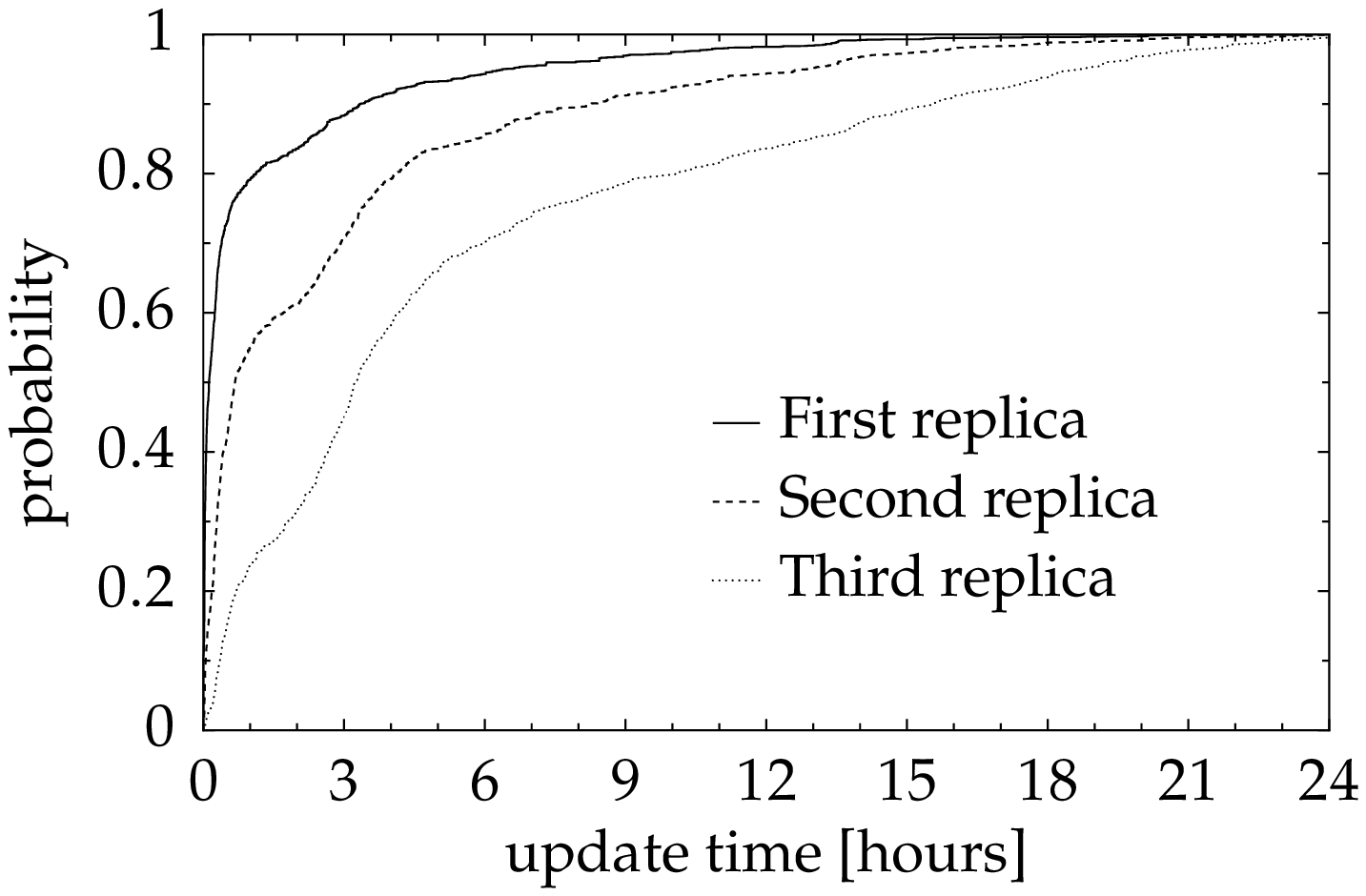}
\caption{The cumulative distribution function for the time of creating a replica of data chunk. The time is relative to the data owner. Lab; data collected over 4 weeks of running time.}
\label{fig:updateTime}
\end{minipage}
\hspace{0.5cm}
\begin{minipage}[b]{0.3\linewidth}
\centering
\includegraphics[width=\textwidth]{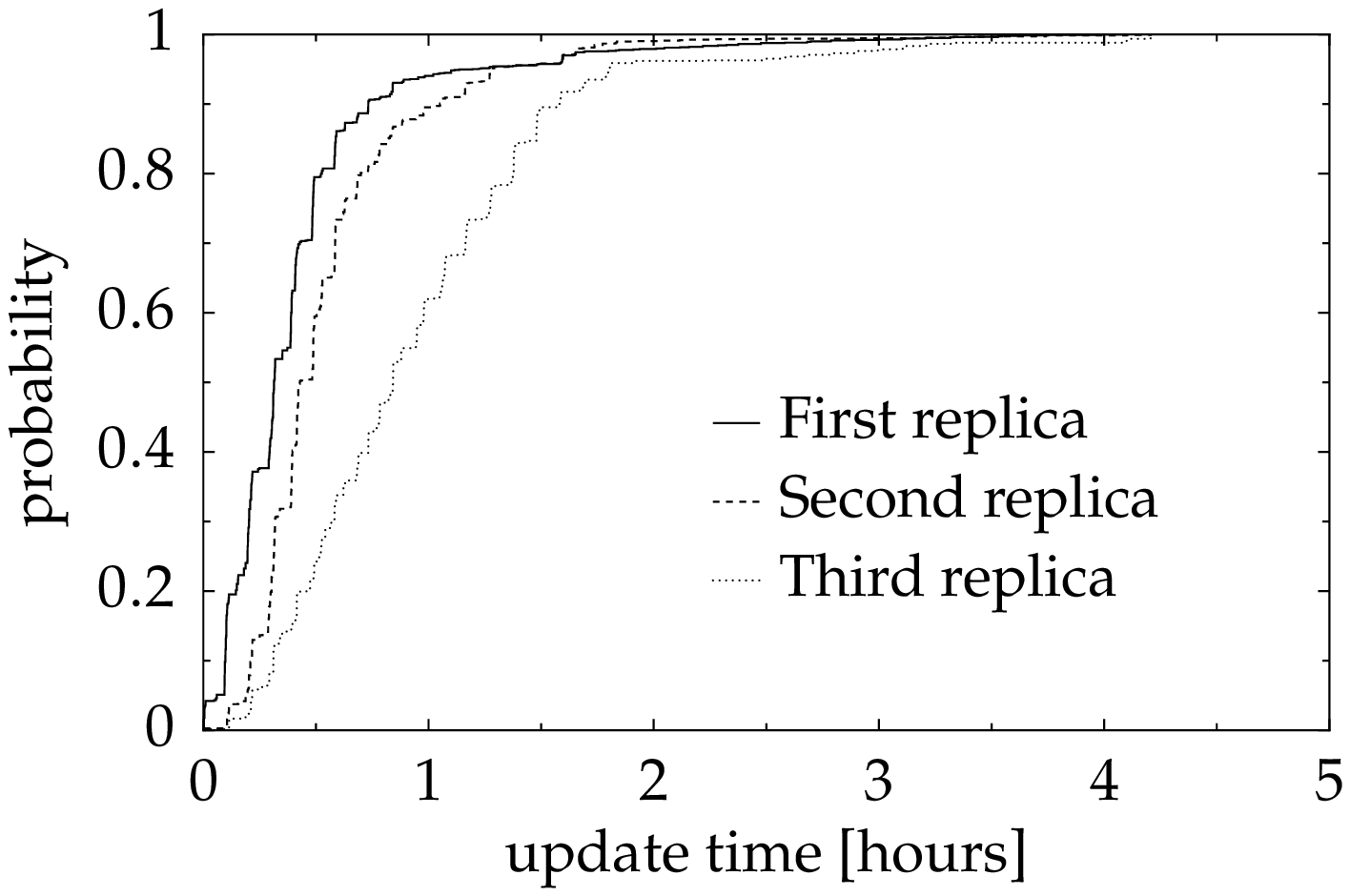}
\caption{The cumulative distribution function for the time of creating a replica. The time is relative to the data owner. Planet-Lab; data collected over 3 days.}
\label{fig:updateTime2}
\end{minipage}
\end{figure*}

The computers in students lab have very low average availability (the median is equal to 13\%). Figure \ref{fig:labAvailability} presents the distribution of the availabilities of the computers in lab. Figure \ref{fig:labUpTime} presents the distribution of the up time of the computers and the time between their consecutive availability periods within a single day (the nights are filtered out). Low availability coupled with long session times constitute a worst-case scenario for a backup application: in contrast to short, frequent sessions, here machines are rather switched on for a day, then switched off when the lab closes and remaining off during the next week.

\subsubsection{PlanetLab}

The experiments on PlanetLab were using 50 machines scattered around Europe. Each machine was provided 10GB of storage space and had 1GB of local data intended to be backed up. The machines were almost continuously available (availability equal 0.91).

\subsection{Asynchronous messages}\label{sec::asynchMesExperiments}

In this subsection we present how the asynchronous messaging influence message delivery time and the probability that the message is delivered. For our analyzes we used the traces of availability from the students computer lab. We varied the number of synchro-peers per peer between 0 and 30. For each number of synchro-peers, we generated 100,000 messages with random source, destination and sent time. Figures \ref{fig:deliveryTime} and \ref{fig:deliveryProbability} present averages and standard deviations.

Figure \ref{fig:deliveryTime} presents the dependency between the number of synchro-peers and the delivery time of the message. Because the message delivery can be accomplished only when the receiver is active, we present delivery time measured from the first availability of the receiver after the message was sent. Ideally, the message should be delivered just after the receiver becomes online. The presented result show that the delivery time, measured from the perspective of the receiver, decreases significantly when using synchro-peers. Additionally, the standard deviation, which because of very low peers availabilities is very high, decreases even more. For the number of synchro-peers higher than 5, the advantage of using more of them becomes less significant. Taking into account that the higher number of synchro-peers results in higher number of messages required for synchronization we decided to use 5 synchro-peers for our prototype system.

Figure \ref{fig:deliveryProbability} presents the dependency between the number of synchro-peers and the probability of a successful delivery a message to any synchro-peer. We are interested in calculating such probability because a message delivered to a synchro-peer is, in fact, a replica of the original message. Thus, synchro-peers should enable message delivery even in case of long term absence of the sender (e.g. caused by non-transient failure). The results show that synchro-peers significantly increase this probability -- with 5 synchro-peers the system delivers 90\% of the messages, while without synchro-peers, more than half of the messages are lost.

\subsection{Replica placement}

\subsubsection{Students computer lab}
The goal of tests on the labs was to verify how the system copes with low availability of the machines. 
For each machine $i$, we set the bandwidth $B_i$ to the same value and the backup window $Des(T_b)$ to 0. As all the machines are in the same local network, there is no geographical distribution of the data. Thus, the utility function (Eq.~\ref{eq::utility}) degrades to the number of replicas and the backup duration (Eq.~\ref{eq::backup-duration}).
This means we wanted to minimize the maximal time required for transferring a data chunk, which means minimizing the load on maximally loaded machine. As the result we expected the machines to be loaded proportionally to their availabilities. By Eq.~\ref{eq::backup-duration}, the load on the machine is proportional to the size of data it replicates; thus for each machines, the total size of replicated data should be proportional to machine's availability. Additionally, storage constraints should influence the amount of data stored.

During the first 3 days of experiments we measured the ratio: total size of data replicated by a peer (in MB) to the peer availability. For each day, we considered only the peers were switched on at least once. We also restricted the measurements only to peers with at least 9GB storage space (that could accommodate, on the average, 3 replicas), to separate the effect of insufficient storage space.
The average values and the standard deviations of the ratios for the 3 days are presented in Table \ref{tab::replicaPlacement}. The standard deviation is low in comparison to the average (they deviations are 18\%, 13\% and 8\% of the corresponding average) which shows that the replicas were distributed according to our expectations.

\begin{table}[tb]
    \centering
    \caption{The ratio: total size of replicated data (in MB) to the availability for the first 3 days of experiments in the lab environment.}
    \label{tab::replicaPlacement}
    \begin{tabular}{ | l | l | l |}
    \hline
    & \multicolumn{2}{c|}{\textbf{Utility (weighted replicated data)}} \\ \hline
    \textbf{day} & \textbf{average} & \textbf{standard deviation} \\ \hline \hline
    1 & 34487 &  6086 \\ \hline
    2 & 60489 &  8141 \\ \hline
    3 & 69658 &  5496 \\ \hline
    \end{tabular}
\end{table}

\subsubsection{PlanetLab}
The goal of PlanetLab tests was to verify how our system handles geographic distribution and heterogeneity of the machines.
In this environment we required the far replica to have TTL distance from the owner in range $\langle 3, 8 \rangle$ ($remote_{min} = 3$ and $remote_{max} = 8$) and other replicas to be as close to the owner as possible ($close_{max} = 0$). 
Additionally we set the bandwidth $B_i$ to 500 KB/s for half of the machines, and 1000 KB/s for the other half. We also set the backup window $Des(T_b)$ to 4500s. 
Each machine had the same amount of local data (1GB); the disk space limit was 4GB.
We expected that the low-bandwidth machines will be less loaded than those from the high-bandwidth group. Assuming that machines are continuously available, a low-bandwidth machine should replicate at most 2.25GB; and a high-bandwidth machines at most 4.5GB.

We tested two parameter settings that differed by the weight assigned to geographical distribution of replicas (see Section \ref{sec::utility-function}). For $M = 1$ (which means increasing the backup duration of a single chunk by 1s is equally unwanted as increasing the TTL distance of this chunk by 1), the average TTL distance between the replica and the owner was equal to 11.6 (std dev. 3.7). In this case only two machines exceeded their backup window (by at most 108 s). For $M = 0.01$ (which means increasing the backup duration of a single chunk by 100s is as bad as increasing the TTL distance of a single chunk by 1), the replicas had better geographic distribution: mean TTL equals 8.1 (std dev. 3.8). However, the backup duration was increased -- 13 machines exceeded their backup window. The average excess of the backup window was equal to 222s (5\% of the backup window) and the maximal 415s (9\% of the backup window).

\subsection{Duration of backup of a data chunk}
We measured the time needed to achieve the consecutive redundancy levels (the number of replicas) for each data chunk. The time is measured relative to the data chunk owner online time: we multiplied the absolute time by the owner's availability. We consider the relative time as a more fair measure because: (i) the transfer to at least the first replica requires the owner to be available; (ii) data can be modified (and thus, the amount of data for backup grows) only when the owner is available; (iii) we are able to directly compare results from machines having different availabilities.

The distribution of time needed to achieve the consecutive redundancy levels is presented in Figure~\ref{fig:updateTime} (lab) and Figure~\ref{fig:updateTime2} (PlanetLab).

\subsubsection{Lab}

The average time of creating the first, the second and the third replica of a chunk are equal to, respectively, 1.1h, 2.7h and 5.5h (the average time needed to create any replica is equal to 3.1h). We consider these values to be satisfactory as the average relative time for transferring a single asynchronous message holding no data (message with 0 synchro-peers), calculated based on availability traces, is equal to 2.6h. 

The maximal values, though, are higher: 24h, 29h and 32h. These high durations of replication are almost entirely the consequence of peers' unavailability. The maximal time needed to deliver an asynchronous message with 3 synchro-peers is of the same order (21.5h, measured relatively to source online time, see Section~\ref{sec::asynchMesExperiments}). Moreover, if we measure only the nodes with more than 20\% average availability, the times needed to create the replicas are equal to 1h, 1.6h and 3h and maximal values are equal to 12h, 18h and 20h.

\subsubsection{PlanetLab}

The average times needed for creating the first, the second and the third replica are equal to, respectively, 0.5h, 0.7h and 1.1h. The maximal values are equal to 4.0h, 4.2h, and 4.2h. These values are significantly better than in case of the students computer lab even though the distance between the machines is much higher and the computers in students lab are connected with a fast local network. This once again proves that the unavailability of the machines is the dominating factor influencing the backup duration.

The average time needed for transferring a data chunk is equal to 0.76h. Assuming that the transfer times of chunks are similar, if the chunks are transferred sequentially then the transfer of the half of data is finished within 0.76h. If the chunks are transferred concurrently then almost all the chunks are transferred within 0.76h. Having 1GB of local data and 3 replicas in both cases we can assume that 1.5GB of data is transferred within 0.76h. This gives an estimated throughput of 4.49Mb/s (Planet-Lab uses standard Internet connections).

\section{Conclusions}
We present an architecture of a p2p backup system based on pair-wise replication contracts. 
In contrast to storing the data in a DHT, in our approach the placement can be optimized to a specific network topology, which allows to take into account e.g. geographical dispersion of the nodes.

We have implemented a prototype and tested it on 150 computers in our faculty and 50 computers in PlanetLab. 

During implementation and initial tests we encountered numerous issues we did not expect: e.g., updating data catalog remotely whenever any contract is changed is highly inefficient; revoking the contracts cannot be done asynchronously; changing contracts too often is inefficient; each contract must be kept by both the data owner and the replicator and the two versions have to be kept consistent. We think that these problems should motivate others to verify their ideas, in addition to simulations, by constructing prototype implementations.

Our most important result is that the backup time increases significantly if machines are weakly-available: from 0.76h for nearly always-available Planet-Lab nodes to 3.1h for our lab with just 13\% average availability. This \emph{cost of unavailability} makes some environments less suitable for p2p backup. The irregular environments negatively influence the maximal durations of data transfer. Choosing machines with better availability strongly reduces this effect (for instance, by restricting our lab environment to machines with more than 20\% availability, the average backup time decreases from 3.1h to 1.9h). Moreover, in enterprise environments such irregular availabilities should not be the case. There, however, the machines may have their specific, regular availability patterns. In such case it may be valuable to use more sophisticated availability models.

Yet, as our main conclusion we must stress that it is possible to build an efficient and reliable backup system, even on the environment with weakly-available machines having irregular session times.
It is possible to take advantage of the heterogeneity of the p2p environment, in particular: the geographic dispersion of the machines, the network connections between the machines; we can use machines with different bandwidths, disk spaces and availabilities. We built a scientific prototype and we managed to run it on 150 machines --- these results are promising and might be considered as the proof of the concept for designing the full efficient and reliable p2p backup system.

\bibliographystyle{abbrv}
\bibliography{bapap}

\end{document}